\documentclass[aps,prb,twocolumn,showpacs,twoside]{revtex4}
\usepackage{graphicx}\usepackage{psfrag}
\newcommand{\mybf}[1]{{\bf #1}}
\renewcommand{\vec}{\mybf}

\usepackage{graphicx} \usepackage{amssymb}
\usepackage{fancybox}\usepackage{subfigure}\usepackage{psfrag}
\newcommand{\GR}{G_\mathrm{R}}\newcommand{\GA}{G_\mathrm{A}}

\newcommand{\insfig}[2]{\begin{minipage}[c]{#1}\includegraphics[width=#1]{#2}
\end{minipage}}
\newcommand{\insfigh}[3]{\begin{minipage}[c]{#1}\includegraphics[height=#2]{#3}
\end{minipage}}

\begin{document}

\title{Electron kinetics in isolated mesoscopic rings
driven out of equilibrium}
\author{V.I.Yudson$^{1,2}$}
\affiliation{$^1$ Center for Frontier Science, Chiba University,
1-33 Yayoi-cho, Inage-ku, Chiba 263-8522, Japan\\
$^2$Institute of Spectroscopy, Russian Academy
of Sciences, 142090, Troitsk, Moscow region, Russia}
\author{ V.E Kravtsov$^{3,4}$}
\affiliation {$^3$International Centre for Theoretical Physics, P.O. Box
586, 34100 Trieste, Italy \\
$^4$Landau Institute for Theoretical Physics, Russian
Academy of Sciences, 2 Kosygina str., 117940, Moscow, Russia}
\date{\today}
\begin{abstract}
Kinetic theory of nonlinear current response to an external field is
developed for mesoscopic normal metal rings threaded by a magnetic flux.
General expressions for direct current (DC) are derived for a
non-equilibrium regime. These expressions describe simultaneously
a contribution to DC made by a non-equilibrium (external) field, as well
as the contribution caused by interaction with intrinsic fields.
Contributions of electron-electron and electron-phonon
interactions to the direct current in a non-equilibrium systems are
estimated. The kinetic equation for electrons in a pumped disordered
metal ring is solved with taking into account the Coulomb and
electron-phonon interactions. This gives an estimate of an overheating
of the pumped electron system.
\end{abstract}
\draft
\maketitle
\section{Introduction}
\label{sec:intro}

Usual objects for experimental and theoretical studies of mesoscopic
phenomena in linear or non-linear conductivity are systems with two
or more external leads. Because of experimental
difficulties, mesoscopic systems without leads have received
less attention, except for a long-standing puzzle of persistent
currents (PC) in normal metal rings in the Aharonov-Bohm configuration
\cite{PC,PCE1,PCE2,MW,Jar,Rab,Mon1}. As the persistent
current is an equilibrium mesoscopic phenomenon, most of theoretical
papers deal with the thermodynamical properties of the electron system;
it is assumed usually that the thermodynamic equilibrium has been
sustained in the experimental investigations cited.

Although an early theoretical study \cite{KY} of non-equilibrium
conductivity in mesoscopic rings has predicted a peculiar effect of a
direct current (DC), which is induced by an external alternating (ac)
electromagnetic field, there has not been performed experimental
search for this effect until very recently \cite{Rab}.

The interest to non-equilibrium mesoscopic phenomena in systems
without leads has increased recently in connection with a
hypothesis \cite{M,KA} that the observed magnitudes of PC may include
a large DC contribution, induced by an uncontrolled external
noise field. It has been supposed also \cite{G,M} that this
uncontrolled noise may be the reason of the anomalous electron
dephasing rate discovered in experimental studies
\cite{MW} of weak localization corrections to the conductance of long
wires. Comparing the dephasing rate caused by an external
electromagnetic field \cite{AAK} in long wires, and the
magnitude of DC current \cite{KY} induced in a ring by the external
field with the same parameters (both quantities are linear in the
external field intensity), a universal relationship between the two
quantities has been proposed recently \cite{KA}. However, the situation
in this field remains still unclear and requires further research, both
experimental and theoretical.

This set of problems together with the arising experimental
interest, necessitate a deeper study of quantum kinetical processes
in systems without external leads. The present paper is devoted
to this study. An important physical difference between systems
with and without leads is that in systems with leads, one may
consider electron transport through the system even neglecting
processes of inelastic scattering of electrons. The role of
a ``thermostat" may be played by scattering processes of electrons
within the leads or in the external part of the electric circuit.
On the contrary, in systems without leads (like metal rings) the
meaning of a steady-state response to an external field is not well
defined unless dissipation processes are taken explicitly into account.

In an earlier paper \cite{YKK} we have considered these features
of small mesoscopic systems in the situation where the main channel
of the energy dissipation is provided by the electron exchange with
neighboring reservoirs, so that the system allows for a simplified
"dynamical" description in terms of the free electron model.
Here we develop a self-consistent kinetic theory of nonlinear current
response to an external field in mesoscopic normal metal rings
threaded by a magnetic flux. The theory takes into account inelastic
electron scattering processes, first of all the processes
of the electron-phonon interaction, which are responsible for
establishing a steady-state regime in the excited electron system.
With the use of the Keldysh technique we have derived a generic
expression for the direct (zero-frequency) current, which includes
terms of two kinds: ``kinetic" terms vanishing in the equilibrium
and "thermodynamic" terms, which describe the interaction induced
correction to PC. These two groups of terms have different analytical
structure, that corresponds to the physical difference between
retarded nature of the response current and equilibrium processes;
the latter allow for a weighted combination of emission-absorption
processes without causality requirement for ``absorption to precede
emission" \cite{KYcm}.

This approach provides a regularized description of the system.
Moreover, it allows to resolve problems connected with anomalous
``diffusion'' modes of zero frequency and momentum (in ref.\cite{YKK}
we call them ``loose diffusons''): we show that for the
distribution function obeying the steady-state kinetic equation,
the contribution of ``zero-diffusion'' modes vanishes identically
(see Appendix). However, this modes are important in transient regimes.

In the presence of an external field the difference
between PC and DC parts of the zero-frequency current has a
somewhat conventional meaning. This is due to the abovementioned
strong non-linearity of the electron system, which restricts
validity of the formal expansion into powers of the external field.
In fact, the PC part of the zero-frequency current, defined
formally as proportional to the zero-th power of the
external field, is modified as compared to its equilibrium value.
This occurs due to a field-induced change of the steady-state
distribution function of electrons. Such an ``incoherent" kinetic
effect is an unavoidable consequence of the external field action.

A steady-state regime in systems without leads may only be achieved
via a mechanism of energy transfer from the pumped
electron system to an external thermostat. As is usually assumed,
this transfer is provided by electron-phonon interaction. The
efficiency of this mechanism decreases with decreasing temperature
that may lead to an overheating of the electron system
(note, that more efficient inelastic electron-electron collisions
conserve energy of electrons and therefore cannot provide cooling
of electron system).

Study of field-induced modifications of the state of electron system
is the major task of the present paper. In Sec. II we consider in
detail the kinetic equation (KE) for electrons excited by an external
field and cooled via interaction with a phonon bath. We have found a
solution to this KE in the limit of relatively frequent
inelastic electron-electron collisions. In this adiabatic limit
the state of electron system may be considered as a
quasi-equilibrium one, described by the Fermi distribution function
with an effective temperature. This state relaxes slowly to a
steady-state regime via electron-phonon inelastic scattering
processes. We have found an explicit expression for the steady
state electron temperature.  We analyze also to what extent
the absence of a global equilibrium between electron and phonon
systems may modify the value of the direct current.

The paper has the following structure.
General consideration of equilibrium and non-equilibrium direct
current is carried out in section II. Model description, notations and
formulation of the problem are given in the subsection II A. Formal
expressions for various contributions to the direct current are
derived by means of the Keldysh technique in subsection II B.

In section III we study a field-induced non-equilibrium energy
distribution of electrons.
Kinetic equation for electrons with taking into account electron-electron
and electron-phonon interaction is described in subsection III A.
A steady state solution to this equation in the adiabatic regime
and the electron temperature relaxation rate are given
in the subsection III B.
A field-induced overheating of electrons and its physical consequences
are discussed in subsection III C.

In section IV we estimate additional contributions to direct current
caused by the absence of a complete equilibrium between electrons
and phonons (subsection IV A) and by a deviation of electron
distribution function from the equilibrium one (subsection IV B).
We show that the latter contribution vanishes in the
constant density of states approximation (for systems with electron-hole
symmetry) while the former one is quite small. These study
justifies the applicability of a free electron model for
description of DC induced by an external field of a moderate intensity.

In section V we summarize the obtained results.

\section{Equilibrium and non-equilibrium direct current}

\subsection{A model and formulation of the problem}
\label{sec:Model}

We consider a normal metal ring of circumference $L$ and cross-section
area $S$, threaded by a magnetic flux $\varphi$. The ring is assumed
to be thin, i.e. $\sqrt{S} \ll L$. It is assumed also that $L \gg l$,
where $l$ is the electron mean free path with respect to elastic
scattering by impurities.
For a time independent magnetic flux $\varphi$, the ring is known
to possess an equilibrium persistent current $I_{PC}$.
A time-dependent magnetic flux would lead to the direct (zero-frequency)
current along the metal ring, this current is a nonlinear response
to the external perturbation \cite{KY}.

Before specifying a particular mechanism of the external action, we
emphasize a basic feature of the response problem in systems without
external leads. A steady-state regime in such systems may be reached
only due to
a balance between an incoming energy flux, pumped to the system, and an
outgoing energy flux to a surrounding thermostat. Therefore, the
mechanism of the energy dissipation in systems without leads is vitally
important and should be explicitly taken into account; the response problem
becomes a kinetical one. This is in contrast to systems with
leads where the energy balance may be provided by outgoing electrons
and the system allows for a dynamical description (see discussion
in \cite{YKK}).

Below we shall consider a kinetical problem of the system response to
an external field with taking into account energy dissipation processes.
We assume that the ``cooling'' of electrons is provided by
interaction with phonons which transfer efficiently the excess energy to
the surrounding thermostat (note that the electron-electron interaction
itself does not change the total energy of the electron system).
As to the field that excites the electron system, the analysis we present
below may be applied for various physical mechanisms.

In particular, in order to keep the connection with our previous
studies \cite{KY,KA,YKK,Y}, we shall consider the action of an
external electromagnetic field with the component
${\cal E}(t) = -(1/c)(d/dt)A(t)$ along the circumference. Such
a field may arise, for instance, if the magnetic flux contains a time
dependent part. We represent the vector potential $A(t)$ in the form:
\begin{eqnarray}\label{A}
A(t) = \int^{\infty}_0 d\omega [A(\omega)\exp{(-i\omega t)} + c.c]
\end{eqnarray}
and assume that this external field has the Gaussian statistics
with the spectral correlation function $S_A(\omega)$ determined by
\begin{eqnarray}\label{SA}
< A(\omega)A^*(\omega')> = S_A(\omega)\delta(\omega - \omega').
\end{eqnarray}
In case of a narrow spectral width (quasi-monochromatic field)
the above expressions reduce to
$A(\omega) = A_0 \delta(\omega - \omega_0)$ and
$S_A(\omega) = |A_0|^2\delta(\omega - \omega_0)$, respectively.

We are interested in the response to the perturbation
\begin{eqnarray}\label{V}
V(t) = -(e/c)A(t)\int d{\bf r}\Psi^{\dagger}({\bf r})\hat{{\bf v}}
\Psi({\bf r})\, ,
\end{eqnarray}
where $\Psi^{\dagger}({\bf r})$ is the electron annihilation
operator and $\hat{{\bf v}} = - i\nabla/m - eA/c$ is the electron velocity
operator ($m$ is the electron mass). More specifically, we are interested
in a field-induced electron current of zero frequency (i.e. a direct current)
flowing along the ring.

In addition to this particular mechanism of the external action,
we shall analyze also a more general situation, assuming only that
due to some ``external perturbation" there arises a non-equilibrium
steady-state in electron-phonon system of the sample, and we shall
study how this influences the direct current. A physical origin of
the perturbation may be also purely ``intrinsic", caused for instance
by the energy release in course of a slow relaxation of the
non-equilibrium nuclear system of the sample.
In this connection we also mention the slow ortho-para conversion of
the $H_2$ molecules inside a metal. The corresponding heat release
of 1 $ppm$ of $H_2$ in $Cu$ at 4 $K$ could be as large as $5\times
10^{-3}$ $nW/g$ even weeks after a cooling-down. The similar heat
release is observed from the quartz glass \cite{Pobell}.

\subsection{Direct current, general expressions}
\label{sec:DC}

First we shall discuss general properties of the direct current in
the Aharonov-Bohm geometry. We shall use the Keldysh formalism which
allows to take simultaneously into account interactions with external and
internal fields (see, e.g., a review \cite{RS}). The current flowing along
the ring is given by
\begin{eqnarray}\label{I}
I(t) = -i\frac{e}{2L}\mbox{\bf Tr}\{\hat{v}_xG^K(t,t)\},
\end{eqnarray}
where $e = -|e|$ is the electron charge,
$\hat{v}_x = -i(\nabla_x - eA/c)/m$ is the electron velocity operator
($x$-axis is chosen along the circumference), and
$G^K \equiv (\underline{\hat{G}})_{12}$ is the Keldysh component of
the Keldysh matrix Green's function in the standard triangular
representation \cite{RS}, and the trace is taken over spin and space
coordinates. In a steady state of the electron system in the absence
of interaction and time dependent external fields,
the Fourier transform of $G^K(t-t')$ is given by
\begin{eqnarray}\label{GK0}
G^{K}_{0}(E) = f_{E}[G^R_{0}(E)-G^A_{0}(E)] \,\, ,
\end{eqnarray}
where $G^{R(A)}_{0}(E)$ is the retarded (advanced) Green's function,
and $f_{E} = 1 - 2n_{F}(E)$ is connected with the electron energy
distribution function $n_F(E)$. If the considered steady state corresponds
to the thermal equilibrium with temperature $T$, $f_{E} = \tanh{E/(2T)}$.
An equilibrium persistent current for a non-interacting electron system
is given by
\begin{eqnarray}\label{PC}
I_{0}= \frac{e}{L}\int \frac{dE}{4\pi i} \, f_{E} \mbox{\bf Tr}\{\hat{v}_x
\left[G^{R}_{0}(E)-G^{A}_{0}(E)\right]\} \, .
\end{eqnarray}
In the presence of an interaction or an external perturbation, we
have instead of Eq.(\ref{GK0})
\begin{eqnarray}\label{GK}
G^K(E) = \left(\underline{\hat{G}}_0 +
\underline{\hat{G}}_0 \,\underline{\hat{\Sigma}}\, \underline{\hat{G}}_0 + ...
\right)_{12}
\end{eqnarray}
where $\underline{\hat{\Sigma}}$ is a self-energy matrix.
In the minimal (second) order of the perturbation theory, the
self-energy splits into independent parts
$\underline{\hat{\Sigma}} = \underline{\hat{\Sigma}}_{f}
+ \underline{\hat{\Sigma}}_{e-ph} + \underline{\hat{\Sigma}}_{e-e}$,
which correspond to different physical perturbation mechanisms
(an external field, electron-phonon and electron-electron interactions,
respectively). Corresponding contributions to the
direct current $I = \overline{I(t)}$ have very similar structure, so
it is sufficient to write down explicitly only expressions for the
case of electron-phonon interaction.

\subsubsection{Electron-phonon interaction}

The time-independent current may be
represented in the form:
\begin{equation}\label{3I}
I = \overline{I(t)} = I_0 + I_{1} + I_{2} + I_{3}.
\end{equation}
Here
\begin{eqnarray}\label{I1}
&&I_{1}=\frac{e}{4L}\int \frac{dE \, d\omega}{(2\pi)^2}\,
\mbox{\bf Tr} \{\hat{v}_x
\left[G^{R}(E)\hat{g}G^{A}(E-\omega)\hat{g}G^{A}(E) - \right. \nonumber\\
&& \left. G^{R}(E)\hat{g}G^{R}(E-\omega)\hat{g}G^{A}(E)\right]
\left([f_{E} - f_{E - \omega}]\hat{D}^K(\omega) \right.
 \nonumber \\
&& \left.  +
[ f_{E}f_{E - \omega} - 1] \Delta\hat{D}(\omega)\right) \} \, ,
\end{eqnarray}
where $\hat{g}$ is an operator vertex of the electron-phonon
interaction $\hat{g}\Psi^{\dagger}\Psi {\bf u}$ that will be specified
later, ${\bf u} = \{u_{\alpha}\}$ is the displacement vector;
$G^{R,A}$ are electron Green's functions in the absence of
interactions (for simplicity, here and below we omit the corresponding
subscript ``$0$'');
$\Delta\hat{D} \equiv \hat{D}^{R} - \hat{D}^{A}$;
$\hat{D}^{R(A)}$ and $\hat{D}^{K)}$ are components of the phonon
matrix Green's function $\underline{\hat{D}}$:
$D^K_{\alpha,\beta}(1,1') =
-i<\left[ u_{\alpha}(1), \, u_{\beta}(1')\right]_{+}>$, etc.
In the short-hand notation the integrand in Eq.(\ref{I1}) may
be represented as $\mbox{\bf Tr}\{{\bf J}_1\}$, where ${\bf J}_1$
has the following structure
\begin{eqnarray}\label{I1sh}
{\bf J}_{1} &=& ({\bf RAA}-{\bf RRA})\,
[(f-f')D^K \nonumber \\
&+& (ff'- 1)\Delta D],
\end{eqnarray}
The parts $I_{2}$ and $I_{3}$ have the same integral representation as
Eq.(\ref{I1}) but with different trace structures instead of
Eq.(\ref{I1sh}):
\begin{eqnarray}\label{I2sh}
{\bf J}_{2}&=&[{\bf RAR}D^R + {\bf ARA}D^A](1 - ff') \, ;
\end{eqnarray}
\begin{eqnarray}\label{I3sh}
{\bf J}_{3}&=&{\bf RRR}[fD^K + (ff'- 1)D^{R}] \nonumber \\
&-& {\bf AAA}[fD^K - (ff'- 1)D^{A}] \, .
\end{eqnarray}
Below we shall treat the phonon subsystem as a large
reservoir in equilibrium. In this case
\begin{eqnarray}\label{DK}
\hat{D}^K(\omega) = N(\omega)\left[\hat{D}^R(\omega) -
\hat{D}^A(\omega)\right]  \, ;
\end{eqnarray}
$N(\omega) = 1 + 2n_B(\omega)$, where $n_B(\omega)$ is the
boson (phonon) occupation number; in the thermal equilibrium
the function $N(\omega) = \coth{[\omega/(2T)]}$.
The part $I_{1}$ (\ref{I1}) describes the contribution to
the direct current that, being averaged over disorder, is not
exponentially small at high frequencies of bosons
$\omega\gg E_{c}$, where $E_c = D/L^2$ is the Thouless energy,
$D$ is the diffusion coefficient of electrons.
This is because the combinations
${\bf R}(E-\omega){\bf A}(E){\bf R}(E)$ and ${\bf A}(E-\omega)
{\bf A}(E){\bf R}(E)$ allow to build a zero-frequency
cooperon \cite{KY}. The part $I_{2}$
(\ref{I2sh}) is a boson analog of the Ambegaokar-Eckern
contribution \cite{AE} to the averaged persistent current that
is exponentially small at high frequencies of bosons.
Finally, the part $I_{3}$ does not contribute to the disorder
averaged current but exhibits only in mesoscopic fluctuations.

A beauty of the expressions is that the ``kinetic" part
$I_{1}$ is identically zero in the complete equilibrium.
Indeed, using (\ref{DK}) we can
represent Eq.(\ref{I1sh}) in the form:
\begin{eqnarray}\label{I1mod}
{\bf J}_{1} &=& ({\bf RAA}-{\bf RRA})\,\Delta D\times
\\ \nonumber & &[(f-f')N(\omega) + ff'- 1],
\end{eqnarray}
where the last square bracket vanishes in the complete equilibrium.
The existence of the kinetic contribution Eq.(\ref{I1mod})
is the main and generic difference between a non-equilibrium
steady state and the complete thermodynamic equilibrium.

\subsubsection{Electron-electron interaction}

Expressions for $I_{1}$-$I_{3}$ are determined by
Eqs.(\ref{I1})-(\ref{I3sh}) with substitutions:
$\hat{g} \rightarrow 1$ and $\underline{D}(\omega)
\rightarrow \underline{V}(\omega)$, where
$\underline{V}$ is the Keldysh matrix for the screened
Coulomb interaction, $V^{R(A)} = V_0/[1 - V_0\pi^{R(A)}]$
and $V^{K)} = V^R \pi^{K} V^A$; here $V_0$ is the usual
(unscreened) Coulomb potential and $\underline{\pi}$ is
a (matrix) polarization operator \cite{RS} (see section IV.B).

\subsubsection{Interaction with an external field}

The contribution to the direct current made by an
external classical field does not contain a "thermodynamical"
part Eq.(\ref{I2sh}). This is due to general (causal) analytical
properties of the response: there may be only zero or one
change of the analyticity ($R \leftrightarrow A$) between
the beginning and end points of the electron lines
(see e.g. \cite{YKK} and references therein).
Expressions for the field-induced direct current are given by
Eqs.(\ref{I1}), (\ref{I1sh}), and (\ref{I3sh}) with the following
modifications: $\hat{g} \rightarrow -e\hat{v}_x/c$,
\begin{eqnarray}\label{DKS}
\hat{D}^K(\omega) \rightarrow - 2\pi i S_{A}(\omega) \, ,
\end{eqnarray}
and omission of terms proportional to $(ff' - 1)D^{R(A)}$.

\subsubsection{External field vs. non-equilibrium nodes}

One can easily see from Eqs.(\ref{DK}) and (\ref{DKS}) that the
effect of an external field is similar to the action of a
non-equilibrium part of the phonon field described
by a non-equilibrium component
of the phonon energy distribution function
$N_{neq}(\omega) = N(\omega) - \coth{[\omega/(2T)]}$.
We arrive at an important conclusion: the absence of a complete
equilibrium in the electron-phonon system of a mesoscopic ring
may result in a non-equilibrium direct current similar to the
DC induced by an external classical field.

In the equilibrium, the sum of $I_{2}+ I_{3}$
gives the quantity $\sigma_{eq}$ from Ref.\cite{KYcm} (if we use the
relationship $D^{R}(\omega)=-D^{A}(\omega)$
used in Ref.\cite{KYcm}).

\section{Non-equilibrium energy distribution in a pumped electron system}

\subsection{Kinetic equation}
From Dyson equations $\left(\underline{\hat{G}}^{-1}_0 -
\underline{\hat{\Sigma}}\right)\underline{\hat{G}} = \hat{I}$ and
$\underline{\hat{G}}\left(\underline{\hat{G}}^{-1}_0 -
\underline{\hat{\Sigma}}\right) = \hat{I}$ for the matrix Green's
function it follows
\begin{eqnarray}\label{DE}
\left[\underline{\hat{G}}^{-1}_0\,,\,\,\underline{\hat{G}}\right] =
\left[\underline{\hat{\Sigma}}\,,\,\,\underline{\hat{G}}\right] \, ,
\end{eqnarray}
where $\underline{\hat{G}}^{-1}_0(1,1') = [ i\partial_{t_1} - H({\bf r}_1)]
\delta( 1-1')$. Taking coordinate trace from the Keldysh
component of the above equation we obtain:
\begin{eqnarray}\label{DKE}
2\pi\nu {\cal V} \partial_{t} f_{E}(t) = \mbox{\bf Tr}\{\Delta \Sigma G^K -
\Sigma^K\Delta G\} \, ,
\end{eqnarray}
where $\nu$ is the averaged electron density of states at the Fermi-level,
${\cal V}$ is the sample volume,
$\Delta \Sigma = \Sigma^R - \Sigma^A$, and the
averaging over disorder realizations is implied on the right hand
side. Using the second order expression for
$\underline{\hat{\Sigma}}$ we arrive at the kinetic equation in
the standard form
\begin{eqnarray}\label{KE}
\partial_{t} f_{E}(t) = {\cal I}_{f} + {\cal I}_{e-ph} +
{\cal I}_{e-e} \, .
\end{eqnarray}
Here the first term on the right hand side describes pumping
by the external electromagnetic field
\begin{eqnarray}\label{ext}
{\cal I}_{f} &=& \frac{1}{2\tau_{f}}
\int^{\infty}_0 d\omega \tilde{S}_A(\omega)\nonumber \\
&&[f_{E+\omega} + f_{E-\omega} -
2f_{E}] \, ,
\end{eqnarray}
where
\begin{eqnarray}\label{norm}
\tilde{S}_A(\omega) = S_A(\omega)/\int^{\infty}_0 S_A(\omega)d\omega
\end{eqnarray}
is a normalized spectral distribution of the external field and
\begin{eqnarray}\label{tau}
\frac{1}{\tau_{f}} = 2D(e/c)^2\overline{A^2(t)} = 4D(e/c)^2
\int^{\infty}_0 S(\omega) d\omega
\end{eqnarray}
is a field induced dephasing \cite{AAK}.

The second, electron-phonon collision term in Eq.(\ref{KE})
is given by:
\begin{eqnarray}\label{St}
&&{\cal I}_{e-ph} =  \frac{i}{4\pi \nu {\cal V}}
\int^{\infty}_{-\infty}\frac{d\omega}{2\pi} \int d{\bf r}d{\bf r'}
\nonumber\\
&&\{\Delta G(E)_{\bf r, r'}[\hat{g}_{\alpha}\Delta G(E - \omega)
\Delta D_{\alpha,\beta}(\omega)\hat{g}_{\beta}]_{\bf r', r}\}
\nonumber\\
&&\left(N(\omega)[f_{E} - f_{E - \omega}] + f_{E}f_{E-\omega} - 1\right)
\end{eqnarray}

\subsubsection{Model of the electron-phonon interaction}
We assume that the mesoscopic ring is embedded into a medium with
a high thermal conductivity. This allows us to consider the phonon
subsystem of the ring as an equilibrium thermal bath. To simplify
the problem, here we do not take into account effects of phonon
scattering at the interface, the presence of surface (interface)
phonons, etc. The electron interaction with long-wavelength crystal
lattice displacements (acoustic phonons) is treated in the
``jelly model'' and is determined by the standard deformation
potential interaction
\begin{eqnarray}\label{def}
H^{def}_{e-ph} = \frac{iC}{\sqrt{\cal V}}
\sum_{\bf p,q} c^{\dagger}({\bf p} + {\bf q})c({\bf p}) ({\bf q u_{q}}) \, ,
\end{eqnarray}
where the deformation potential $C = p_{F}v_{F}/3$, $c^{\dagger}$ and $c$
are the electron creation and annihilation operators, and ${\bf u_{q}}$
is the Fourier transform
\begin{eqnarray}\label{uq}
{\bf u_q} = \frac{1}{\sqrt{\cal V}}\int d{\bf r} \exp{(-i{\bf qr})}
{\bf u}({\bf r}).
\end{eqnarray}
In systems with electron scattering by impurities (i.e., in real
metals) one has to take into account also the effect of phonon-induced
impurity displacements (see Refs.\cite{Sch,RS87}).
This results in an additional part of
the electron-phonon interaction that does not have free parameters once
the electron-impurity interaction is chosen
\begin{eqnarray}\label{imp}
H^{imp}_{e-ph} = -\frac{i}{\sqrt{{\cal V}}}
\sum_{\bf p,k,q} U({\bf k})c^{\dagger}({\bf p} + {\bf q} + {\bf k})
c({\bf p}) ({\bf k u_{q}}) \, .
\end{eqnarray}
Here
\begin{eqnarray}\label{Uk}
U({\bf k}) = (1/{\cal V}) \sum_{a} U_{a}({\bf k})
\exp{(-i{\bf kr}_{a})} \, ,
\end{eqnarray}
where
\begin{eqnarray}\label{Ua}
U_a({\bf k}) = \int d {\bf r} U_{a}({\bf r}) \exp{(-i{\bf kr})}
\end{eqnarray}
is the Fourier transform of the potential of the a-th impurity.
For the sake of completeness we write down also an expansion of
the displacement ${\bf u(r)}$ over the phonon creation and
annihilation operators
\begin{eqnarray}\label{ua}
u_{\alpha}(r,t) &=& \frac{1}{\sqrt{\rho_m {\cal V}}}
\sum_{{\bf q},j} e^{(j)}_{\alpha}\sqrt{\frac{1}{2\omega_j(q)}}\nonumber \\
&&\left( a_j(q)\exp{[i({\bf qr}-\omega_j(q)t)]} + \mbox{H.c.} \right)
\end{eqnarray}
and the expression for the phonon Green's function
\begin{eqnarray}\label{DD}
&&\Delta D_{\alpha,\beta}(1,1') = \nonumber \\
&&\frac{1}{{\cal V}}\sum_{{\bf q},j}\int \frac{d\omega}{2\pi}
\Delta D^{(j)}_{\alpha,\beta}({\bf q},\omega)
\exp{[i({\bf qr}-\omega t)]} \, ;
\end{eqnarray}
\begin{eqnarray}\label{Dq}
&&\Delta D^{(j)}_{\alpha,\beta}({\bf q},\omega) = 
-\frac{\pi}{\rho_m\omega_j(q)}\times \\ \nonumber & &\left[ \delta(\omega -
\omega_j(q))
- \delta(\omega + \omega_j(q))\right]\eta^{(j)}_{\alpha,\beta}({\bf q}) .
\end{eqnarray}
In Eqs.(\ref{ua})-(\ref{Dq}) $\rho_m$ is the material density;
$j$ accounts for the longitudinal ($j=l$) and two (equivalent)
transverse phonon polarizations ($j=t_1$ and
$j=t_2$). For longitudinal phonons $\eta^{(l)}_{\alpha,\beta}({\bf q}) =
q_{\alpha}q_{\beta}/q^2$; while for transverse phonons
\begin{eqnarray}\label{tr}
\sum_{j = t_1, t_2} \eta^{(j)}_{\alpha,\beta} =
\delta_{\alpha,\beta} - q_{\alpha}q_{\beta}/q^2 \, .
\end{eqnarray}
To perform the disorder averaging of the expression
$\mbox{\bf Tr}\{\Delta G(E)\hat{g}_{\alpha}\Delta G(E - \omega)
\hat{g}_{\beta}\}$ in the collision integral (\ref{St}),
one should take into account that the impurity potential $U$ enters
also the interaction vertex $g^{imp}_{\alpha} = -iU({\bf k})k_{\alpha}$,
that corresponds to Eq.(\ref{imp}). Various contributions to the averaged
value of $\mbox{\bf Tr}\{\Delta G(E)\hat{g}_{\alpha}\Delta G(E - \omega)
\hat{g}_{\beta}\}$ are represented by diagrams shown in Fig.1
(they are similar to those describing a disorder averaged electron
self-energy in the paper by Reizer and Sergeev\cite{RS87}).

\begin{figure}
\begin{minipage}[h]{8cm}
\subfigure[]{\includegraphics[width=1.5cm]{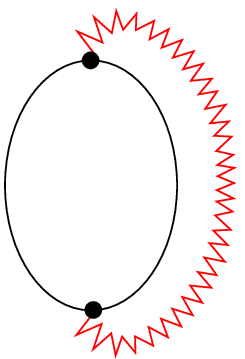}}\hspace{.3cm}
\subfigure[]{\includegraphics[width=1.5cm]{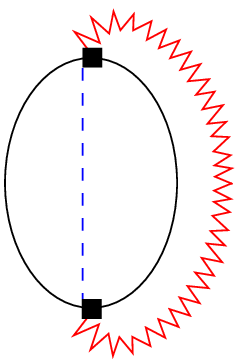}}\hspace{.3cm}
\subfigure[]{\includegraphics[width=1.5cm]{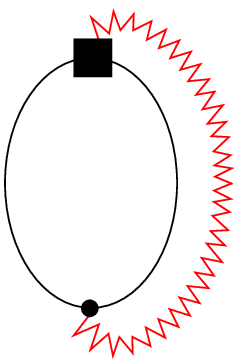}}\hspace{.3cm}
\subfigure[]{\includegraphics[width=1.5cm]{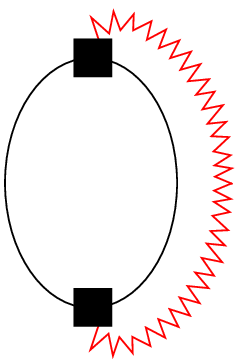}}
\end{minipage}
\begin{minipage}[b]{8cm}
\subfigure[]{\includegraphics[width=1.5cm]{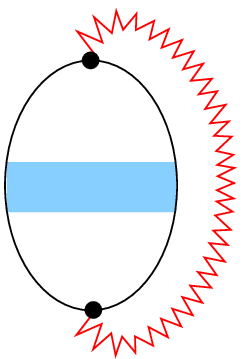}}\hspace{.3cm}
\subfigure[]{\includegraphics[width=1.5cm]{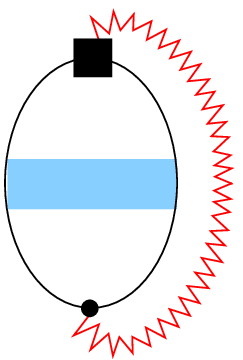}}\hspace{.3cm}
\subfigure[]{\includegraphics[width=1.5cm]{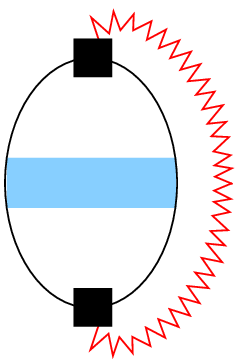}}
\end{minipage}
\begin{minipage}[b]{8cm}
\subfigure[An effective (renormalized) vertices of
the electron-phonon-impurity interaction.]{%
$\insfig{2cm}{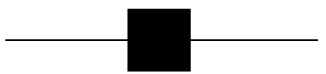}=\insfig{2cm}{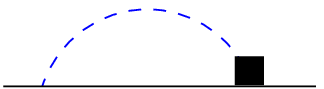}+\insfig{2cm}{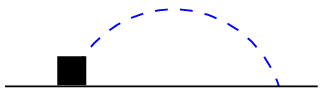}$}
\end{minipage}
\begin{minipage}[b]{8cm}
\subfigure[Diffuson
proparators.]{$\insfigh{2.5cm}{1cm}{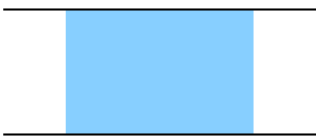}=\insfigh{1cm}{1cm}{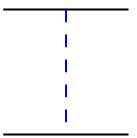}+\insfigh{2cm}
{1cm}{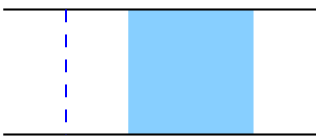}$}
\end{minipage}
\caption{(a-g) - diagrams for the averaged kernel of the
electron-phonon
collision integral in the kinetic equations. Lines describe
electron propagators; dashed and zigzag lines correspond to the
correlation function of the disorder potential and to the
phonon propagator,
respectively. Circles and small squares denote vertices of the
usual (deformation potential) electron-phonon interaction and of
the electron-phonon-impurity interaction, respectively; large
squares denote an effective (renormalized) vertices of
the electron-phonon-impurity interaction, determined by the diagram h.
\label{fig1}}
\end{figure}

First we consider the contribution of transverse acoustic phonons.
The deformation part Eq.(\ref{def}) of the electron-phonon
interaction is absent in this case, as well as the diagrams
a, c, e, and f of Fig.1. Moreover, there is no effect of
``diffuson dressing" of interaction vertices (the diagram g):
a dressed vertex with an entering phonon wave vector ${\bf q}$ is
proportional to ${\bf q}$ and does not contribute due to the
transverse structure Eq.(\ref{tr}) of the phonon Green's function.
Thus, the leading contribution is given by simple (``Drude'')
diagrams (b and d) and one obtains the following
expression for the ${\cal I}_{e-ph}$ (\ref{St})
\begin{eqnarray}\label{Stf}
&&{\cal I}_{e-ph} = \int^{\infty}_{-\infty}d\omega K(\omega)
\left(N(\omega)[f_{E+\omega} + f_{E-\omega} \right.
\nonumber \\
&& \left. - 2f_{E}] + f_{E}[f_{E+\omega} - f_{E-\omega}] + 2\right) \, .
\end{eqnarray}
Here the kernel $K(\omega)$ is given by
\begin{eqnarray}\label{K}
K(\omega) = \frac{3\beta_t\omega}{2p^2_Flv_t}
\Phi_t\left(\frac{|\omega|l}{v_t}\right) \, ,
\end{eqnarray}
where $v_t$ is the sound velocity of transverse acoustic phonons
and a dimensionless (transverse) coupling constant $\beta_t$ is
determined as
\begin{eqnarray}\label{beta}
\beta_t = \left(\frac{p_Fv_F}{3}\right)^2\frac{\nu}{\rho_mv^2_t} \,
\end{eqnarray}
and the function $\Phi_t(x)$ is defined by
\begin{eqnarray}\label{Pt}
\Phi_t(x) = 1 + 3[x - (x^2 + 1)\mbox{arctan}(x)]/(2x^3).
\end{eqnarray}
In the limiting cases: $\Phi_t(x) \approx x^2/5$ at $x \ll 1$ and
$\Phi_t(x) \approx 1$ at $x \gg 1$ .
The same function arises in studies of phonon effects on the
conductivity of disordered metals \cite{RS87,Pti}.

Similarly, one may obtain a contribution of longitudinal phonons.
Now all the diagrams a-g in Fig.1 should be taken into account.
In the limit of small phonon momenta $q \ll 1/l$, each of the
diagrams (e, f, and g) with the diffuson installation is
considerably larger than the corresponding generic diagram (a, c,
and d, respectively). However, in the leading order in $1/(ql)^2$
the diffuson diagrams mutually cancel and their remaining parts
are comparable with contributions made by diagrams a-d.
Such a suppression of the diffuson dressing of electron-phonon
interaction vertices, demonstrated by Reizer and Sergeev \cite{RS87},
occurs due to a destructive interference of contributions made by
the usual deformation potential Eq.(\ref{def}) and by its
electron-phonon-impurity counterpart Eq.(\ref{imp}): the diagrams
e and f are canceled by the cross-terms described by the diagram
f and its symmetric partner (not shown). The physical mechanism of
such a cancellation is connected with the Galilean transformation
from the laboratory system to a system moving together with the
crystal lattice\cite{Sch,RS87}.

The resulting expressions for the contribution of longitudinal
phonons differ from Eq.(\ref{K})-(\ref{Pt}) by the replacement
$\beta_t \rightarrow
\beta_l$ (where $v_t \rightarrow v_l$) and $\Phi_t(x)
\rightarrow \Phi_l(x)$, where $\Phi_l(x) \sim x^2$ at $x \ll 1$.
Typically, $v_t$ is by factor 2-3 smaller than $v_l$, that allows
one to neglect safely the contribution of longitudinal phonons
\cite{Pti}.

\subsection{Steady state solution to kinetic equation}

We are interested in a steady-state regime of the electron subsystem
subjected to a stationary external pumping. The only collision
channel that takes off the injected energy, is the electron-phonon
interaction. The electron-electron interaction conserves the total
energy but leads to a redistribution of electrons over energy levels.
In a closed electron system with a given energy this would
effectively lead to a quasi-equilibrium Fermi distribution with
some effective temperature $T_e$. We shall assume that this process
of establishing a quasi-thermal distribution in the electron system
is faster than the cooling rate due to the electron-phonon collisions
(similar approach was used in Ref.\cite{Spi} in studying
non-equilibrium currents in a metal ring).
This allows us to use an adiabatic approach: we put
$f_{E}(t) = \mbox{tanh}[E/(2T_e(t))]$ and $N(E) =
\mbox{coth}[E/(2T)]$ into Eq.(\ref{KE}) and integrate over
$E$ with the weight $E$. Due to the energy conservation,
the electron-electron collision term makes no contribution and we
arrive at a closed equation for the effective electron temperature
$T_e$:
\begin{eqnarray}\label{Te}
&&\frac{4\pi^2T_e}{3}\frac{dT_e}{dt} = \frac{\omega^2_{0}}{\tau_{f}}
\\ \nonumber
&+& 4\int^{\infty}_{0}d\omega \,
\omega^2K(\omega)\left[\mbox{coth}\left(\frac{\omega}{2T}\right)
- \mbox{coth}\left(\frac{\omega}{2T_e}\right)\right].
\end{eqnarray}
where $\omega_0$ is the "mean-square-root" of the external field
frequency
\begin{eqnarray}\label{o0}
\omega^2_0 \equiv \int^{\infty}_{0} \, \omega^2\tilde{S}(\omega) \,
d\omega \, .
\end{eqnarray}
In the derivation of Eq.(\ref{Te}) we have used an identity
$ff'-1 = - N(E-E')(f-f')$ for the Bose and Fermi distribution functions
at the same temperature.

An effective electron temperature $T_e$ in a steady-state regime
is determined by the stationary solution to Eq.(\ref{Te}). Small
deviations $\delta T_e$ from the stationary solution $T_e$
are described by a linearized Eq.(\ref{Te}):
\begin{eqnarray}\label{dTe}
d\delta T_e/dt = - \delta T_e/\tau_T,
\end{eqnarray}
where the temperature relaxation rate is given by
\begin{eqnarray}\label{tT}
\frac{1}{\tau_T} = \frac{3}{2\pi^2 T^3_e}
\int^{\infty}_{0} \, \frac{\omega^3K(\omega)}
{\mbox{sinh}^2[(\omega/(2T_e)]} d\omega \, .
\end{eqnarray}
To obtain explicit expressions for $T_e$ and $\tau_T$, we
restrict our consideration to the limiting cases where
the electron elastic mean-free path $l$ is small or large as
compared to the phonon wavelength $\sim v_t/\omega$ which
corresponds to typical transfer energies $\omega \sim T_e$.

Using Eqs.(\ref{K}) and (\ref{Pt}) in the case of {\it small}
$x = \omega l/v_t$, we arrive at the following expression
for the effective steady-state electron temperature $T_e$
\begin{eqnarray}\label{Tes}
T_e = \left[ T^{6} + \frac{\hbar^4\omega^2_{0}p^2_Fv^3_t}
{288\zeta(6)\beta_tk^6_Bl\tau_f}\right]^{1/6} \,
\end{eqnarray}
and for the temperature relaxation rate $1/\tau_T$:
\begin{eqnarray}\label{tTs}
\frac{1}{\tau_T} = \frac{54\zeta(6)\beta_tl(k_BT_e)^4}
{5\pi^2 \hbar^2p^2_Fv^3_t} \,.
\end{eqnarray}
In these expressions $\zeta(x)$ is the Riemann zeta-function
($\zeta(6) \approx 1$); for further estimates we have written
explicitly the Planck and Boltzmann constants, $\hbar$ and $k_B$.
As we see from Eq.(\ref{Tes}), a noticeable overheating
takes place at intensities of the external field determined
by Eq.(\ref{tau}) and the condition
\begin{eqnarray}\label{taus}
\frac{1}{\tau_f} \sim \frac{3\cdot10^2 \beta_tl (k_BT)^{6}}
{\hbar^4\omega^2_{0}p^2_Fv^3_t} \, .
\end{eqnarray}
At relatively high temperatures $T$ or at a considerable
overheating ($T_e \gg T$), the assumption of small ratio
$x = \omega l/v_t \sim T_el/v_t$ may be violated.
For the opposite case of relatively {\it large} transfer
energy $x = \omega l/v_t \sim T_el/v_t \gg 1$ the kernel
$K(\omega) \sim \omega$ and we find:
\begin{eqnarray}\label{Tel}
T_e = \left[ T^{4} + \frac{\hbar^2\omega^2_{0}p^2_Fv_tl}
{72\zeta(4)\beta_tk^4_B\tau_f}\right]^{1/4} \,
\end{eqnarray}
and
\begin{eqnarray}\label{tTl}
\frac{1}{\tau_T} = \frac{216\zeta(4)\beta_t(k_BT_e)^2}
{\pi^2 p^2_Fv_tl} \,.
\end{eqnarray}
For the validity of the used adiabatic approach we should
require the smallness of the temperature relaxation rate
$1/\tau_T$ as compared to the rate $1/\tau_{e-e}$ of
establishing a quasi-thermal Fermi distribution due to
electron-electron collisions. An estimate for $\tau_{e-e}$
follows from the electron-electron collision term \cite{RS}
in the considered quasi-1D geometry:
\begin{eqnarray}\label{tee}
\frac{1}{\tau_{e-e}} \sim \frac{k_BT_e}{\hbar g}\sqrt{E_c/E_{cut}} \, .
\end{eqnarray}
Here $g = E_c/\Delta$ is the dimensionless conductance ($\Delta =
1/(\nu {\cal V})$) is the mean electron energy level spacing),
$E_{cut}$ is a typical low energy cutoff that depends on the
relationship between $E_c$ and $T_e$, but at any case $E_{cut}\tau
\ll 1$, where $\tau = l/v_F$ is a mean free time for elastic scattering.
Comparing the electron ``thermalization'' rate $1/\tau_{e-e}$
with the temperature relaxation rate $1/\tau_T$ Eq.(\ref{tTs})
(considering, for instance, the case of small typical phonon
momenta $k_BT_e/(\hbar v_t) < 1/l$), we obtain:
\begin{eqnarray}\label{ads}
\frac{\tau_{e-e}}{\tau_T} \sim \frac{S}{l^2}
\left(\frac{k_BT_el}{\hbar v_t}\right)^3
\sqrt{E_{cut}\tau} \, .
\end{eqnarray}
For typical experimental configurations the ring cross section $S$ is
comparable with $l^2$, and the ratio $\tau_{e-e}/\tau_T$ is small
due to two other factors on the right hand side of Eq.(\ref{ads}).
This justifies the validity of the adiabatic approach.

\subsection{Numerical estimates}

To get an idea about an overheating effect, consider a gold
ring. Following ref.\cite{Pti}, we take
the following material parameters: $\beta = 1.4$, $p_F =
1.3\cdot 10^{-19}\mbox{g}\cdot \mbox{cm}/\mbox{s}$, and
$v_t = 1.2\cdot 10^5 \mbox{cm}/\mbox{s}$.
Then Eq.(\ref{Tes}) may be represented in the form
\begin{eqnarray}\label{est1}
T_e \approx 70 \mbox{mK} \frac{(\omega_{0}\cdot 10^{-9}\mbox{s})^{1/3}}
{[(l\cdot 10^5 \mbox{cm}^{-1})
(\tau_f\cdot 10^8 \mbox{s}^{-1})]^{1/6}} \,
\end{eqnarray}
where it has been assumed that $T_e \gg T$. Eq.(\ref{Tes}) has been
derived under the assumption of small transferred momenta $T_e/v_t \lesssim
1/l$, that may be represented as
\begin{eqnarray}\label{est2}
T_e \lesssim 90 \mbox{mK}/(l\cdot 10^5 \mbox{cm}^{-1}).
\end{eqnarray}
If the fraction on the right hand side of Eq.(\ref{est1}) does not
exceed unity, the expressions (\ref{est1}) and (\ref{est2}) are
compatible.

We see, that the presence of a high frequency external pump field
may result in a noticeable overheating of the electron system.
For $\omega_0 \sim 10^9 \mbox{s}^{-1}$ and $l \sim 10^{-5}$cm, the
electron temperature $T_e$ is measured by tens mK even for such
an external field that leads to a quite moderate field-induced
dephasing rate on a scale of $10^7$-$10^8$s$^{-1}$.
We note also that $T_e$ depends very weakly (as $P^{1/6}$) on the 
pumping power $P$. 

The overheating may put serious obstacles in investigations of
coherent nonlinear phenomena in closed mesoscopic samples.

\section{Non-equilibrium effects on the direct current}

As discussed in the preceding section, pumping the electron
system may result in a noticeable deviation of the electron
distribution function from the equilibrium one. The absence
of the complete equilibrium between the electron and phonon
subsystems may lead to an additional contribution to the
averaged direct current. This contribution cannot be
considered in the framework of a simplified dynamical
treatment used in the earlier work \cite{KY} (see also
\cite{YKK}), it requires the kinetic approach.
Here we shall study this effect. In the first and second
subsections we shall consider non-equilibrium phonon and
Coulomb contributions to disorder averaged DC.
Of principal importance is the non-equilibrium
contribution connected with the part $I_1$ Eq.(\ref{I1})
of the DC, as this part vanishes in the equilibrium.
Calculation of disorder averaged quantities is performed
within the usual formalism of weak localization theory.
The only point to be commented is the appearance of a
diffuson carrying zero frequency and wave vector. This
singular diffuson (``loose'' diffuson \cite{YKK}) is a
typical feature of non-equilibrium problems in mesoscopic
systems.
Such a diffuson may be built by drawing parallel disorder
lines embracing the ``self-energy'' part between R and A
Green's function (see Fig.3 in Appendix). The role of
the anomalous diffuson has been studied earlier \cite{YKK}
in the framework of the dynamical consideration and it is
closely connected with the renormalization of the electron
energy distribution function. Here (see Appendix) we prove
an important identity for the anomalous diffuson. Namely,
we show that when taking into account all the mechanisms
contributing to the ``self-energy'', the diagrams with
the anomalous diffuson cancel each other at a {\it steady-state}
regime.
Thus the loose diffusons signal on the incorrect or
time-dependent energy distribution function, and they cancel
out as soon as the correct steady-state distribution function
$f_{E}$ is substituted in Eq.(\ref{GK0}).

\subsection{Non-equilibrium electron-phonon interaction
effects on DC}

The question about a DC induced due to the absence of the
equilibrium between electrons and phonons, has been put forward
by Spivak, {\it et al.} \cite{Spi}, who have calculated the
DC variance. Here we shall consider not the variance but
the disorder averaged DC (in the presence of a magnetic flux
pierced the ring). As in the preceding sections we restrict
our consideration to the case of transverse acoustic phonons.

The leading contribution to DC is given by a diagram
(see Fig.2) containing a loop of two cooperon propagators
connecting two ``Hikami boxes''(a rhomb and a triangle).
\begin{figure}[h]
\begin{minipage}[b]{8cm}
\psfrag{k+q,0}{$C(0,\vec k)$}\psfrag{q,-w}{$D(\omega,\vec q)$}\psfrag{k,-w}
{$C(\omega,\vec k-\vec q)$}\psfrag{q,w}{}\psfrag{w}{}\psfrag{-w}{}\psfrag{0}{}
\psfrag{R, E}{}\psfrag{A, E}{}\psfrag{R, E-w}{}
\includegraphics[height=4cm]{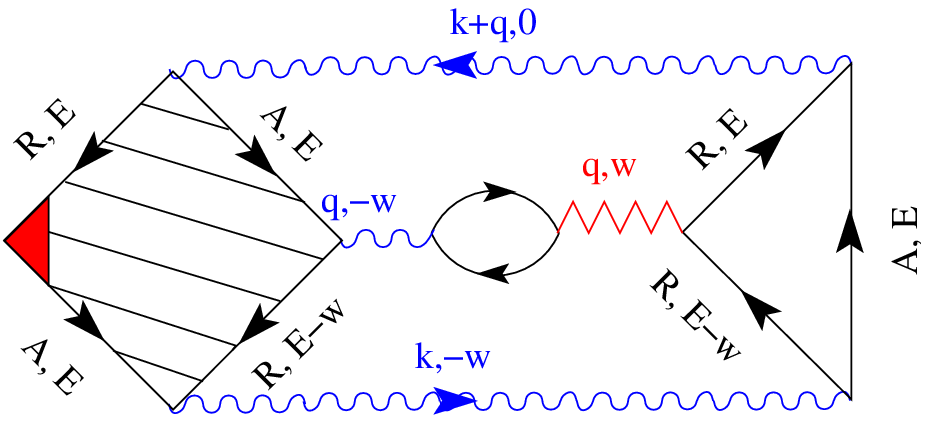}
\end{minipage}\\[1cm]
\begin{minipage}[b]{8cm}
\psfrag{c}{$C$}\psfrag{d}{$D$}
$\insfigh{1.5cm}{0.7cm}{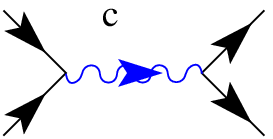}=\insfigh{1.5cm}{0.7cm}{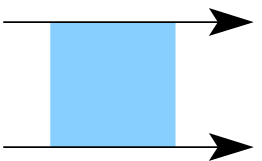},\quad
\insfigh{1.5cm}{0.7cm}{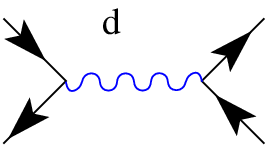}=\insfigh{1.5cm}{0.7cm}{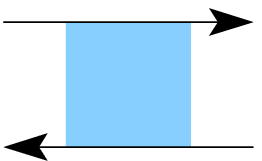}.$
\end{minipage}
\caption{A leading contribution of the electron-phonon or
electron-electron
interaction to a non-equilibrium part of the time-independent (DC)
current. Wavy lines correspond to cooperons (C) or diffusons (D); the zigzag line
corresponds to the screened electron-electron interaction or the phonon propagator.
In the case of longitudinal phonons both
electron-phonon and renormalized electron-phonon-impurity vertices should be taken
into account. For the case of transverse phonons the diffuson should be
omitted, and the rhomb and triangle are directly connected by the phonon
propagator (the zigzag line) terminated by the renormalized
electron-phonon-impurity vertices. \label{fig2}} \end{figure}

One of the cooperons carries zero frequency and a small
wave vector $k = 2\pi(n - 2\phi)/L$ along the ring
circumference (i.e.,"x"-axis); $\phi$ is a magnetic flux
measured in units of $hc/|e|$. It is the cooperon that
provides the dependence of the effect on the magnetic flux.
The dependence of the second cooperon $C(k-q,\omega)$ on
the magnetic flux is negligible at phonon wave vectors
$q >> 1/L$.
Expressions for the Hikami boxes are given by $F_{\alpha}(q)$ for
the rhomb and by $kF_{\beta}(q)$ for the triangle.
The vector function $F_{\alpha}(q)$ is defined (for the
RRA succession of Green's functions) as
\begin{eqnarray}\label{F}
F_{\alpha}(q) &=& \delta_{\alpha, x}\frac{p_Fl}{3}\frac{1}{V}
\sum_{p}G^R(p+q)G^A(p) \nonumber \\
&+& \frac{i}{V}\sum_{p}v_{x}p_{\alpha}G^R(p+q)[G^A(p)]^2
\end{eqnarray}
For both RRA and RAA cases, the corresponding functions $F_{\alpha}$
coincide and equal to
\begin{eqnarray}\label{Fh}
F_{\alpha} = \delta_{\alpha, x}\frac{2\pi \nu \tau p_Fl}{3}h(ql)
\end{eqnarray}
(a longitudinal part $\sim q_{\alpha}$ has been omitted here), where
\begin{eqnarray}\label{h}
h(x) = \frac{\mbox{arctan}(x)}{x} + \frac{3[\mbox{arctan}(x) - x]}{x^3}
\end{eqnarray}
At large $x$, $h(x) \approx \pi/(2x)$. At small $x$, $h(x) \approx
(4/15)x^2$.
For $m$-th harmonic of the phonon induced averaged current we obtain
\begin{eqnarray}\label{Im}
&&I^{(m)}_1 = \frac{1}{6\pi^3}\beta_t \frac{e}{\nu Du_t}
\int^{\infty}_{0}d\omega \, \omega^2 h^2(\omega l/u_t)\nonumber\\
&& [\mbox{coth}(\omega/(2T)) - \mbox{coth}(\omega/(2T_e))]
\mbox{Im}C(\omega/u_t,\omega) \, .
\end{eqnarray}
The range of small momentum transfer $q = \omega l/u_t \ll 1$
corresponds to the diffusion regime, where $\mbox{Im}C(q,\omega) =
\omega/(\omega^2 + D^2q^4)$. In this case the leading contribution
to the integral stems from the range $u^2_t/(v_Fl) \ll \omega \ll u_t/l$
and we obtain
\begin{eqnarray}\label{Is}
I^{(m)}_1 = \frac{\zeta(4)}{\pi^3}\left(\frac{4}{15}\right)^2
\beta_t \frac{el^4(T^4_e - T^4)}{\nu D^3u_t}
\end{eqnarray}
Therefore we have an estimate (for $T_e - T \gtrsim T$):
\begin{eqnarray}\label{est}
I^{(m)}_1 = eE_c [(T_e/E_c)(\tau T_e)(T_e/E_F)(T_e/\omega_D)],
\end{eqnarray}
where $\omega_D \sim u_tp_F$ is the Debye frequency. The factor
$eE_c$ is a typical (mesoscopic) magnitude of a persistent current
in a single ring, but the expression in the square brackets is
extremely small.

Even smaller quantity arises in the case of large momentum transfer
$q = \omega l/u_t \gg 1$ in Eq.(\ref{Im}), that may take place at
relatively large $T_e$, i.e. $T_e > u_t/l$, the latter quantity is
of the order of 100 mK for $u_t = 10^5$cm/s and $l = 10^{-5}$cm.
In this case we shall use the ballistic expression for the
``cooperon'', that reads:
$\mbox{Im}C(q,\omega) = \omega/(v_Fq)^2$ (at $\omega \ll v_Fq$).
As a result, we shall obtain an expression which is by a factor
$[u_t/(lT_e)]^3 \ll 1$ smaller than Eq.(\ref{est}).

We arrive at the conclusion that the non-equilibrium contribution
of electron-phonon interaction to the flux-dependent part of the
averaged DC is negligible. This is compatible with the smallness
of the photovoltaic current variance found by Spivak, {\it et al.}
\cite{Spi}.

\subsection{Non-equilibrium electron-electron interaction effects
on DC}
As noted in sec.II, effects of the absence of the complete
equilibrium in pumped electron systems may result in a non-zero
contribution $I_1$ to DC. This effect takes place for the phonon
contribution to $I_1$ studied in the preceding section, although
the magnitude of the effect turns out to be small. Considering the
contribution of the electron-electron interaction to $I_1$ we
deal with the modification $\underline{\hat{D}}(\omega)
\rightarrow \underline{V}(\omega)$ of the (symbolic) expression
${\bf J}_1$ Eq.(\ref{I1sh}) (see subsection IIB1):
\begin{eqnarray}\label{I1C}
{\bf J}_{1} &=& ({\bf RAA}-{\bf RRA})\,
[(f-f')V^K \nonumber \\
&+& (ff'- 1)\Delta V] \,
\end{eqnarray}
Here $V^{R(A)} = V_0/[1 - V_0\pi^{R(A)}]$
and $V^{K)} = V^R \pi^{K} V^A$
are elements of the Keldysh matrix $\underline{V}$
for the screened Coulomb interaction; $\Delta V = V^R-V^A$;
$V_0$ is the usual (unscreened) Coulomb potential; and
$\underline{\pi}$ is
a (matrix) polarization operator \cite{RS}. In dirty metals
(in the diffusive electron propagation regime) matrix
components of the screened Coulomb ``potential'' are given
by (see \cite{AA,RS}):
\begin{eqnarray}\label{DV}
\Delta V(q,\omega) = - \frac{2i\omega}{Dq^2\nu},
\end{eqnarray}
and
\begin{eqnarray}\label{VK}
V^K(q,\omega) = - \frac{2i\overline{\omega}}{Dq^2\nu},
\end{eqnarray}
where $\nu$ is the density of states (at the Fermi energy)
corresponding to the effective sample dimension, and the
quantity $\overline{\omega}$ is defined by
\begin{eqnarray}\label{O}
\overline{\omega} = \frac{1}{2}\int^{\infty}_{-\infty}
[1 - f_{E}f_{E-\omega}]\, dE \, .
\end{eqnarray}
To obtain the contribution of the part Eq.(\ref{I1C})
to the disorder averaged DC, we should average the
triangle of electron Green's functions
$({\bf G^R(E)G^A(E-\omega)G^A(E)}-
{\bf G^R(E)G^R(E-\omega)G^A(E)})$ and integrate the
resulting expression over $E$. Formally, the leading
contribution to the kinetic part ${\bf J_1}$ of the averaged
DC would be given by the two-cooperon diagram in Fig.2.
However, we will show that this contribution, as well as
contributions of higher order diagrams, vanish after the
integration over $E$. Due to the presence of
distribution functions with the asymptotic properties
$f_{E} \rightarrow \pm 1$ at $E \rightarrow \pm \infty$,
the integration over $E$ runs in a close vicinity of the
Fermi energy. Here we restrict ourselves to the usual
approximation of the weak localization theory, namely,
we neglect a weak dependence of averaged products of
electron Green's functions on the common part $E$ of
their arguments; we neglect also the energy dependence
of the averaged density of states (DOS). In this approximation
the only $E$-dependent part of the integrand is given
by the square bracket $[ \, ... \, ]$ in Eq.(\ref{I1C})
and we find:
\begin{eqnarray}\label{0}
\int^{\infty}_{-\infty}
[\, ... \, ] \, dE = 2\omega V^K(q,\omega) -
2\overline{\omega} \Delta V(q,\omega) = 0.
\end{eqnarray}
(the latter equality follows from Eqs.(\ref{DV}) and
(\ref{VK})).
This result has been obtained without any assumption
about the form of the distribution function $f_{E}$;
the derivation is based on the identity
\begin{eqnarray}\label{fdE}
\int^{\infty}_{-\infty}
[f_{E} - f_{E - \omega}] dE = 2\omega
\end{eqnarray}
valid for an {\it arbitrary} distribution function
$f_{E}$ (satisfying the standard asymptotic conditions).
We arrive at the conclusion that effects of non-equilibrium
and non-thermal electron distribution do not contribute to
the ``kinetic'' part $I_1$ of the direct current.
Of course, there is always a ``thermodynamic'' part $I_2$
of the direct current, considered by Ambegaokar and Eckern
\cite{AE}, but this part is nonzero even in the complete
equilibrium and is only weakly changed by a relatively weak
pumping.

It should be emphasized once again that the vanishing of the
$I_1$ part of the electron-electron contribution to DC has been
obtained in the approximation of the energy-independent DOS
and the averaged electron triangle, as well as
neglecting disorder-induced correlations between
the ``electron triangle'' and the polarization operator.
Results obtained beyond the constant DOS approximation will
be reported elsewhere \cite{CK}.

\section{Conclusions}

We have considered kinetics of closed mesoscopic samples (metal
rings) subjected to an external ac pump field. To provide
a consistent description of excitation and relaxation processes
in the pumped electron system, electron-phonon interaction has to
be taken explicitly into account. With the use of Keldysh technique
we have derived general expressions for a time-independent electric
current along the ring in the presence of a static magnetic flux.
These expressions describe simultaneously a thermodynamical
persistent current (PC) and a direct current (DC) induced nonlinearly by
the ac field or, in general, by an arbitrary perturbation of the
equilibrium state.
The kinetic approach allows to avoid a problem of anomalous
(loose) diffusons \cite{YKK}. The singularity represented by a
loose diffuson in the particular case of
diffusive mesoscopic systems, is generic to non-equilibrium systems
described in the framework of the Keldysh technique. The origin
of this singularity is the ansatz Eq.(\ref{GK0}). The singularity
arises if the energy distribution function $f_{E}$ does not satisfy
the kinetic equation (e.g., if it is chosen to be the equilibrium
Fermi distribution). We show that the anomalous diagrams with loose
diffusons vanish in the steady state regime if $f_{E}$ is chosen to
make the inelastic collision integral vanish.

Solving the kinetic equation for the electron system, pumped by
an external field and cooled through the phonon bath, we have
derived expressions for an effective electron temperature $T_e$.
We have found that for appreciable external field intensities,
there may be a noticeable overheating of the electron system with
$T_e$ being on the scale of tens mK. This overheating may hinder
manifestations of mesoscopic effects (like PC phenomena).
The analysis restricts the range of permissible external pump
intensities.

In addition to crude thermal effects, there may be more subtle
phenomena. Namely, the DC may be induced not only by the
external field but also due to the absence of a complete equilibrium
between electron and phonon subsystems. To have a self-consistent
description, we have examined such non-equilibrium contributions to
DC. We have found that the phonon contribution to DC caused by
the difference of phonon ($T$) and electron ($T_e$) temperatures,
is negligible as compared to the typical PC magnitude.
We have studied also the contribution to DC caused by the Coulomb
interaction of non-equilibrium electrons, for a non-Fermi
distribution of electrons over energy levels. We have found
that within the framework of usual approximations of the weak
localization theory (neglecting energy dependence of the
electron density of states near the Fermi surface), this
non-equilibrium distribution vanishes identically.
This analysis generalizes, supports, and shows the limits of
applicability of the results of earlier works \cite{KY,KA},
where only the field-induced contribution to DC was considered.

\section*{Acknowledgments}
We are thankful to O.L.Chalaev for useful discussions.
The work was partially supported by
the grant from the Ministry of Science and Education of
Japan (Mombusho), by the RFBR grant No.98-02-16062,
and by the grant ``Nanostructures'' from the Russian
Ministry of Science (V.I.Y.).

\section{Appendix: Vanishing of the
anomalous diffuson contribution}

As mentioned in the text, there may be anomalous diagrams with
a ``loose diffusion'' part \cite{YKK}. Such a diffuson
may be constructed of retarded ($G^A$) and advanced ($G^R$)
Green's functions in- and outgoing from a self-energy
installation $\Sigma$, see Fig.3.
\begin{figure}[h]
\psfrag{r}{$\GR$}\psfrag{a}{$\GA$}\psfrag{s}{$\Sigma$}
\includegraphics[width=5cm]{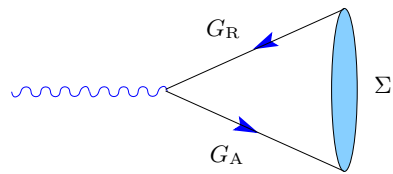}
\caption{A ``head'' of the anomalous diffuson.\label{fig3}}
\end{figure}

As this installation conserves
energy (in case of an external field we separate a part with
equal numbers of absorbed and emitted ``photons''), the energy
arguments of $G^R$ and $G^A$ coincide, hence the loose diffusion
carriers a zero frequency. An anomaly arises in the case, where
the diffusion ``head'' ($G^R\Sigma G^A$) is averaged over the
disorder independently of the averaging the rest part of the
diagram. Due to the restoration of the translational symmetry, the
averaged ``head'' transfers no momentum to the loose diffuson.
Thus, we arrive at a singularity connected with a diffuson of zero
frequency and wave vector.

Our present task is to show that the contribution of anomalous diagrams
vanishes in a steady state regime. Namely, we shall argue for
the vanishing of the diffuson ``head'' ${\cal H}(E)$
\begin{eqnarray}\label{H1}
{\cal H}(E) = \mbox{\bf Tr}\left[\underline{G}(E)\underline{\Sigma}(E)
\underline{G}(E)\right]^K_{RA} \, ,
\end{eqnarray}
where all functions are in Keldysh's space; the trace is performed
over space coordinates; the subscript $RA$ means that we need to keep
only the part with outgoing $G^R$ and incoming $G^A$ lines.
Using the relationship $G^K(E) = f_{E}\Delta G(E)$, we have:
\begin{eqnarray}\label{H2}
{\cal H}(E) &=& \mbox{\bf Tr}\left[G^R(E)
[- f_{E}\Sigma^R(E) \right. \nonumber \\
&+& \left. \Sigma^K(E) + f_{E}\Sigma^A(E)]G^A(E)\right]^{irr} \, .
\end{eqnarray}
Performing here the disorder averaging one should take into account
only an ``irreducible'' part (this is marked by a superscript $irr$),
i.e., those diagrams which do not include an anomalous diffuson ladder
built between $G^R$ and $G^A$ lines, as this loose diffuson has been
already extracted from the ``head'' (see Fig.3).
First, we shall establish a useful identity for an ``irreducible vertex''
${\cal R}^{irr}$
\begin{eqnarray}\label{R1}
{\cal R}^{irr}({\bf r}_1, {\bf r}_2; E) = \int d{\bf r}\,
[G^R({\bf r}, {\bf r}_1; E) G^A({\bf r}_2, {\bf r}; E)]^{irr}
\end{eqnarray}
formed by $G^R$ and $G^A$ lines connected with the same site.
We shall use the following trick. For the moment, we treat an
infinitesimal positive quantity $\delta$ in the definition of
exact electron Green's functions ${\hat G}^{R(A)} =
[E - {\hat H} \pm i\delta/2]^{-1}$ as a finite constant, with
taking the limit $\delta \rightarrow +0$ afterwards. This allows
to deal with a {\it ``reducible vertex''} ${\cal R}$ with no
restriction for the arrangement of disorder lines on averaged diagrams.
This relationship between ${\cal R}$ and ${\cal R}^{irr}$ is shown
symbolically in Fig.4.
\begin{figure}[h]
\psfrag{r}{$\GR$}\psfrag{a}{$\GA$}
$\insfigh{1.2cm}{1.5cm}{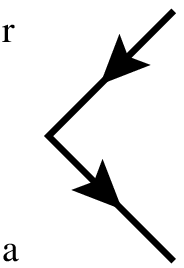}=\insfigh{1.2cm}{1.5cm}{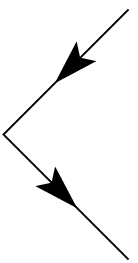}+\insfigh{1.2cm}{1.5cm}
{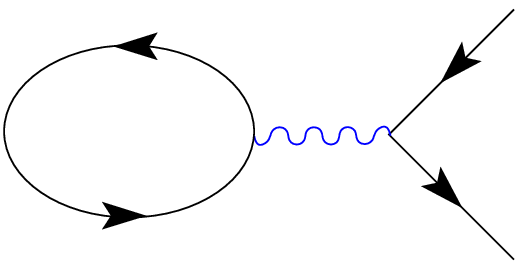}$
\caption{Relationship between reducible (bold) and irreducible
vertices.
\label{fig4}}
\end{figure}

The reducible vertex contains an anomalous part
with a diffusion of zero frequency and momenta. This diffusion equals
$1/\delta$. In the leading order in $\delta\tau(E) \ll 1$, where
$1/\tau(E)$ is the rate of elastic scattering, we obtain
\begin{eqnarray}\label{R2}
{\cal R}^{irr}({\bf r}_1, {\bf r}_2; E) = \delta \tau(E)
{\cal R}({\bf r}_1, {\bf r}_2; E).
\end{eqnarray}
Note that neither the left, nor the right hand sides of Eq.(\ref{R2})
are singular in the limit $\delta \rightarrow +0$. Representing
the Green's functions in terms of exact eigenstates $\psi_{\mu}$ and
eigenenergies $E_{\mu}$
\begin{eqnarray}\label{GRA}
G^{R(A)}({\bf r}, {\bf r'}; E) = \sum_{\mu} \frac{\psi_{\mu}({\bf r})
\psi^*_{\mu}({\bf r'})}{E - E_{\mu} \pm i\delta/2}  \, ,
\end{eqnarray}
we obtain the following expression for the reducible vertex:
\begin{eqnarray}\label{R3}
{\cal R}({\bf r}_1, {\bf r}_2; E) &=& \int d{\bf r}\,
G^R({\bf r}, {\bf r}_1; E) G^A({\bf r}_2, {\bf r}; E) \nonumber \\
&=& (i/\delta) \Delta G({\bf r}_2, {\bf r}_1; E) \, ,
\end{eqnarray}
where $\Delta G = G^R - G^A$.
With the use of Eqs.(\ref{R2}) and (\ref{R3}), we arrive at the
final expression for the irreducible vertex $R^{irr}$ of
our interest:
\begin{eqnarray}\label{R4}
{\cal R}^{irr}({\bf r}_1, {\bf r}_2; E) = i\tau(E)
\Delta G({\bf r}_2, {\bf r}_1; E).
\end{eqnarray}
As neither the left, nor the right hand sides of Eq.(\ref{R4})
are singular, we may safely take the limit $\delta \rightarrow +0$.
Using Eq.(\ref{R4}) we find for the diffuson head ${\cal H}(E)$
\begin{eqnarray}\label{H3}
{\cal H}(E) &=& - i\tau(E)\mbox{\bf Tr}\left[ [f_{E}\Sigma^R (E) -
\Sigma^K(E) \right. \nonumber \\
&-& \left. f_{E}\Sigma^A(E)]\Delta G(E)\right] \,.
\end{eqnarray}
The trace on the right hand side of the last equation coincides
identically with that on the right hand side of the kinetic
equation (\ref{DKE}). We conclude that the contribution of
anomalous diagrams vanishes in a steady state regime,
if the correct steady-state distribution function
$f_{E}$ is used in the ansatz Eq.(\ref{GK0}).


\begin{references}
\bibitem{PC}
M.Buttiker, Y.Imry, and R.Landauer, Phys.Lett. A {\bf 96},
365 (1983).
\bibitem {PCE1} L.P.Levy, G.Dolan, J.Dunsmuir, and H.Bouchiat,
Phys. Rev. Lett. {\bf 64}, 2074 (1990).
\bibitem {PCE2} V.Chandrasekhar, R.A.Webb, M.J.Brady, M.B.Ketchen,
W.J.Gallagher, and A.Kleinsasser, Phys.Rev.Lett. {\bf 67}, 3578 (1991).
\bibitem{MW} P.Mohanty, E.M.Q.Jariwala, and R.A.Webb,
Phys. Rev. Lett. {\bf 78}, 3366 (1997).
\bibitem{Jar} E.M.Q.Jariwala, P.Mohanty, M.B.Ketchen, and R.A.Webb,
Phys.Rev.Lett. {\bf 86}, 1594 (2001).
\bibitem{Rab} W.Rabaud, L.Saminadayar, D.Mailly, K.Hasselbach, A.Benoit,
and B.Etienne, Phys.Rev.Lett. {\bf 86}, 3124 (2001).
\bibitem{Mon1} See, e.g., review:
G.Montambaux, in: {\it Quantum Fluctuations},
edited by S.Reynaud, E.Giacobino, and J.Zinn-Justin
(Elsevier, Amsterdam, 1997).
\bibitem{KY}
V.E.Kravtsov and V.I.Yudson, Phys.Rev.Lett. {\bf 70}, 210 (1993);
A.G.Aronov and V.E.Kravtsov, Phys.Rev. B {\bf 47}, 13409 (1993);
V.E.Kravtsov, Phys.Lett. A {\bf 172}, 452 (1993).
\bibitem{M} P.Mohanty, Ann. Phys.(Leipzig) {\bf 8}, 7 (1999).
\bibitem{KA}
V.E.Kravtsov and B.L.Altshuler, Phys.Rev.Lett. {\bf 84}, 3394 (2000).
\bibitem{G} B.L.Altshuler, M.E.Gershenson, and I.L.Aleiner,
cond-mat/9803125 (unpublished);
Yu.B.Khavin, M.E.Gershenson, and A.L.Bogdanov,
Phys. Rev. Lett. {\bf 81}, 1066 (1998).
\bibitem{AAK}
B.L.Altshuler, A.G.Aronov, and D.E.Khmelnitskii,
J.Phys. C {\bf 15}, 7367 (1982).
\bibitem{YKK}
V.I.Yudson, E.Kanzieper, and V.E.Kravtsov, Phys.Rev. B
{\bf 64}, 045310 (2001)
\bibitem{KYcm} V.E.Kravtsov and V.I.Yudson, cond-mat/9712149 (unpublished).
\bibitem{Y} V.I.Yudson, Phys. Rev. B {\bf 65}, 115309 (2002).
\bibitem{Pobell} F.Pobell {\em Matter and methods at low temperatures},
Springer-Verlag, p.179.
\bibitem{RS}
J.Rammer and H.Smith, Rev.Mod.Phys. {\bf 58}, 323 (1986).
\bibitem{AE}
V.Ambegaokar and U.Eckern, Phys.Rev.Lett. {\bf 65},
381 (1990); Europhys.Lett. {\bf 13}, 733 (1990); U.Eckern, Z.Phys.
B {\bf 82}, 393 (1991); A.Schmid, Phys.Rev.Lett. {\bf 66},
80 (1991); U.Eckern and P.Schwab, Adv. Phys. {\bf 44}, 387 (1995).
\bibitem{Sch} A.Schmid, Z.Phys. {\bf 259}, 421 (1973).
\bibitem{RS87} M.Yu.Reizer and A.V.Sergeev, Zh.Eksp.Teor.Fiz. {\bf 92},
2291 (1987) [Sov.Phys.JETP {\bf 65}(6), 1291 (1987)].
\bibitem{Pti} N.G.Ptitsina, {\it et al.} Phys.Rev. B {\bf 56}, 10089 (1997).
\bibitem{Spi} B.Spivak, F.Zhou, and M.T.Beal Monod,
Phys. Rev. B {\bf 51}, 13226 (1995).
\bibitem{AA} B.L.Altshuler and A.G.Aronov, in {\it Electron-Electron
Interactions in Disordered Systems}, eds. A.L.Efros and M.Pollak
(North-Holland; Amsterdam, Oxford, New York, Tokyo, 1985), p.1.
\bibitem{CK} O.L.Chalaev and V.E.Kravtsov, cond-mat/0204176
(unpublished).
\end{references}
\end{document}